\documentclass{PoS}

\usepackage{amssymb}
\usepackage{amsmath}
\usepackage{amsfonts}
\usepackage{graphicx}

\title{Theory of electric dipole moments and lepton flavour violation}

\ShortTitle{Theory of EDMs and LFV}

\author{\speaker{Martin Jung}\\ 
        Excellence Cluster Universe\\ Technische Universit\"at M\"unchen\\
  Boltzmannstr. 2\\D-85748 Garching, Germany\\
        E-mail: \email{martin.jung@tum.de}}

\abstract{Electric dipole moments and charged-lepton flavour-violating processes are extremely sensitive probes for new
physics, complementary to direct searches as well as flavour-changing processes in the quark sector. Beyond the ``smoking-gun'' feature
of a potential significant measurement, however, it is crucial to understand their implications for new physics models quantitatively. 
The corresponding multi-scale problem of relating the existing high-precision measurements to fundamental parameters can be approached
model-independently to a large extent; however, care must be taken to include the uncertainties from especially nuclear and QCD
calculations properly. 
}

\FullConference{16th International Conference on B-Physics at Frontier Machines\\
        2-6 May 2016\\
        Marseille, France}

\begin{document}

\section{Introduction}
Electric dipole moments (EDMs) and charged-lepton flavour-violating (cLFV) processes provide competitive means to search for new
physics (NP), complementary to strategies like direct searches at hadron colliders, but also to other indirect searches like the
flavour-changing processes investigated at the flavour factories. The exceptional sensitivity is due to the combination of experimental
precision with a tiny Standard Model (SM) background. The smallness of the latter is related to the very specific connection between
flavour and CP violation in the SM,\footnote{EDMs are T,P-odd; their existence implies also CP violation when assuming CPT to be
conserved as we will in this article.} embodied by the Kobayashi-Maskawa mechanism~\cite{Kobayashi}, which is very effective in
suppressing flavour-changing neutral currents (FCNCs) in the quark sector~\cite{Glashow:1970gm}, much more so in in the lepton sector,
but also flavour-conserving neutral currents involving CP violation.\footnote{An exception to the latter statement is provided by
the gluonic operator $\mathcal{O}_{G\tilde G}\propto\epsilon_{\mu\nu\rho\sigma}G^{\mu\nu}G^{\rho\sigma}$, yielding a potentially very
large contribution to hadronic EDMs which is, however, strongly bounded experimentally, constituting the \emph{strong CP problem}. In
the following it is assumed that this issue is resolved by the Peccei-Quinn- or a similar mechanism~\cite{Peccei:1977hh}.}
The remaining SM contributions are many orders of magnitude below the present limits, \emph{e.g.} $d_n^{\rm SM,CKM}\lesssim
10^{-(31-32)}\,e\,{\rm cm}$~\cite{Khriplovich:1981ca,Gavela:1981sk,McKellar:1987tf,Mannel:2012qk} for the neutron EDM.
The observation of an EDM or cLFV with the present experimental precision would therefore clearly constitute a NP signal.

\section{EDMs}

Sakharov's conditions~\cite{Sakharov:1967dj} require the presence of new sources of CP violation to explain the observed baryon
asymmetry of the universe; while this does not \emph{imply} sizable EDMs, they are generally very sensitive to such sources. 
In fact, generic NP scenarios usually yield contributions that are large compared to experimental limits, implying either a high NP
scale or a very specific structure for additional CP-violating contributions, similar to the situation in the flavour-changing sector.
Casting these qualitative statements into reliable bounds on model parameters requires knowledge of their relation to the experimental
observables -- typically (bounds on) frequency shifts obtained for composite systems. The calculation of these relations proceeds via a
series of effective field theories (EFTs), see \emph{e.g.}
Refs.~\cite{Ginges:2003qt,Pospelov:2005pr,Fukuyama:2012np,Engel:2013lsa} for recent reviews and references therein. Importantly, this
approach allows to perform a large part of the analysis model-independently. The calculation of the matrix elements of the corresponding effective operators often include large uncertainties which have to be taken into account, see
Refs.~\cite{Engel:2013lsa,Jung:2013hka} for recent detailed discussions.
Furthermore, in composite systems different contributions can exhibit cancellations; this issue can already be systematically addressed
for paramagnetic systems~\cite{PhysRevA.85.029901,Jung:2013mg}, and in the future potentially also for diamagnetic
ones~\cite{Chupp:2014gka}.

\subsection{Model-independent constraints from EDM measurements}

The available competitive observables, that is, the EDMs of thorium monoxide (ThO) and ytterbium fluoride (YbF)
molecules~\cite{Baron:2013eja,Hudson:2011zz}, thallium (Tl) and mercury (Hg) atoms~\cite{Regan:2002ta,Griffith:2009zz,Graner:2016ses}
and the neutron~\cite{Baker:2006ts,Afach:2015sja} (see also \cite{Serebrov:2013tba}), are related by calculations on the molecular,
atomic, nuclear and QCD levels to the coefficients of an EFT at a hadronic scale (see, \emph{e.g.},
Refs.~\cite{Pospelov:2005pr,Engel:2013lsa,Khriplovich:1997ga}).
The operator basis consists of (colour-)EDM operators $\mathcal{O}^{\gamma,C}_f$, the purely gluonic Weinberg operator $\mathcal{O}_W$
and T- and P-violating four-fermion operators $\mathcal{O}_{ff'}$ without derivatives ($f^{(\prime)}=e,q$, $q=u,d,s$). Since these
calculations do not depend on the NP model under consideration, this Lagrangian is used as the interface between experiment and
high-energy calculations: the latter provide the model-specific expressions for the corresponding Wilson coefficients , 
with at least one more intermediate EFT at the electroweak scale.

\begin{figure}
 \includegraphics[width=6.5cm]{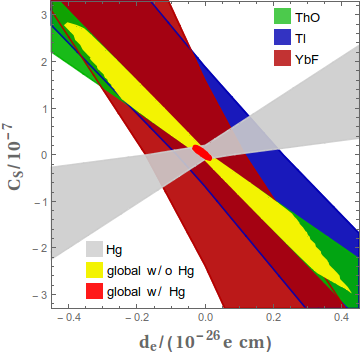}\hfill\includegraphics[width=6.5cm]{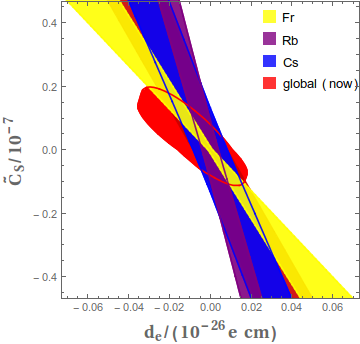}
 \caption{\label{fig::globalFit} Fit to the recent measurements for paramagnetic 
 systems~\cite{Baron:2013eja,Hudson:2011zz,Regan:2002ta}, using additionally the Hg measurement~\cite{Graner:2016ses} (grey band on the
 left). These plots are updated versions of the ones in Refs.~\cite{Jung:2013hka,Jung:2013mg}.
 }
 \end{figure}

In neutral composite systems, the EDMs of the components are shielded; for non-relativistic, point-like constituents this shielding is
perfect, therefore measurements for this type of system rely on the violation of these assumptions~\cite{Schiff:1963zz}. For
paramagnetic systems, relativistic effects can actually lead to \emph{enhancement} factors, if the proton number $Z$ is large
enough~\cite{Sandars:1965xx,Sandars:1966xx,Flambaum:1976vg}, since two contributions scale approximately with $Z^3$: these are the ones
from the electron EDM and the scalar electron-nucleon coupling, $\tilde C_S$.\footnote{Note that $\tilde C_S$ depends in general on the
considered system. However, for the systems at hand (and more generally for heavy paramagnetic systems), it is universal to very good
approximation~\cite{Jung:2013mg}.} Heavy paramagnetic systems can therefore be assumed to be completely dominated by these two
contributions, allowing a model-independent fit to bound and eventually determine \emph{both} contributions, without the assumption of
a vanishing electron-nucleon contribution~\cite{PhysRevA.85.029901,Jung:2013mg}. In practice, there are two complications with this
approach at present, which can however be overcome with additional measurements. Firstly, the ratio of the coefficients of the two
contributions is necessarily similar for heavy paramagnetic systems~\cite{PhysRevA.85.029901}. This problem can be solved by performing
measurements with atoms or molecules with largely different proton numbers, such as rubidium and francium atoms. Lacking such
(competitive) measurements, it is possible to assume \emph{e.g.} the limit from Hg to be saturated by the $d_e, \tilde C_S$
contributions~\cite{Jung:2013mg}:\footnote{Note that we include here the contribution of $d_e$ as well, although its coefficient is
very uncertain~\cite{1402-4896-36-3-011}. We allow for a factor of 2 in this estimate, which is however an arbitrary choice. Additional
calculations are necessary.} this is a conservative procedure, since the EDM of this system is typically dominated by colour-EDM
(cEDM) contributions; the coefficients of the $d_e,\tilde C_S$ contributions in Mercury are about a factor $10^8$ smaller than in
paramagnetic molecules. 
We illustrate this procedure in Fig.~\ref{fig::globalFit}~(left); the fit yields
\begin{equation}
d_e\leq2.7\,10^{-28}e\,{\rm{cm}}\,\,(95\%~{\rm CL})\,,\quad {\mbox{and}}\quad \tilde C_S\leq 1.5\times 10^{-8}\,.
\end{equation} 
These values should be used when extracting bounds on parameters from $d_e$ in any model in which the electron-nucleon contribution
cannot be argued to be negligible.
Furthermore, this 2-dimensional constraint allows to obtain model-independent limits on the EDMs of all other heavy paramagnetic
systems~\cite{Jung:2013mg}. A significant measurement in one of these systems larger than these bounds would indicate an experimental
problem. These limits are orders of magnitude below existing experimental ones.
Importantly, present experiments aim at an even better
sensitivity~\cite{Weissetal,Amini:2007ku,2004APS..DMP.P1056K,Sakemi:2011zz,PhysRevX.2.041009}, as illustrated on the right-hand side
in Fig.~\ref{fig::globalFit}.

The extension of this type of fit to all EDM measurements is clearly possible and has been proposed in Ref.~\cite{Chupp:2014gka}. While
this is complicated by the many potential contributions -- all of the terms mentioned above are relevant in general, this should be
aimed for in the future. Since model-independent bounds/determinations are necessary to determine the specific structure of
CP-violating NP contributions -- and thereby potentially the model itself --, it is essential to have as many measurements in different
systems as possible. An additional complication for the EDMs of diamagnetic systems and neutrons is that the theoretical uncertainties
for the relevant matrix elements are often large and can in some cases preclude the extraction of conservative limits, for instance on
the cEDMs from Hg~\cite{Jung:2013hka}, highlighting the importance of additional theoretical studies, but also further motivating
complementary measurements.

\subsection{\label{sec::EDMsandNP} NP contributions to EDMs}
Reliable limits on parameters in NP models from EDMs are difficult to achieve. Reasons are, apart from the fact discussed previously
that presently less competetive measurements than relevant effective operators exist, the presence of several contributions to each of
these coefficients and the various relevant hierarchies, \emph{i.e.} in mass scales, couplings and loop factors. This
complicates semi-model-independent analyses for classes of models and allows strict statements only under additional assumptions.
However, once a specific model is considered, typically strong correlations exist between EDMs and other CP-violating observables. 

Generic NP contributions at tree- and one-loop level are in conflict with the stringent experimental limits. On the two-loop level,
usually so-called Barr-Zee- and Weinberg diagrams dominate~\cite{Weinberg:1989dx,Dicus:1989va,Barr:1990vd}, which
compensate the additional loop factor by avoiding small mass factors. However, flavour sectors are usually far from generic;
therefore in some cases also tree-level diagrams can be relevant, for example when they involve small mass factors, see below.

In order to demonstrate these qualitative statements in a specific model, we consider a general two-Higgs-doublet model (2HDM).
In this setup, the situation is typically the one described above: four-quark (tree-level) contributions are subleading, one-loop
contributions to (colour-)EDMs are under control (but not necessarily tiny), and two-loop contributions are dominant, but also the
tree-level quark-electron couplings are relevant, despite the small mass factors~\cite{Buras:2010zm,Jung:2013hka}.
To be (even more) specific and able to relate the resulting bounds quantitatively to those from other observables, we will furthermore
consider the Aligned 2HDM (A2HDM)~\cite{Pich:2009sp,Jung:2010ik}, where the Yukawa matrices in each sector are proportional to
each other in order to avoid FCNCs at tree-level, but with complex proportionality factors. 

The electron EDM receives in this class of models contributions mostly from Barr-Zee diagrams; the resulting constraints require
factors at the percent level in addition to the suppression by fermion masses and CKM factors, questioning already the common
assumption that such factors should be $\mathcal O(1)$.
In the A2HDM this can be compared to the absolute value of the same parameter combination obtained from leptonic and semileptonic
decays~\cite{Jung:2010ik,Celis:2012dk}, which is about a factor 1000 weaker, demonstrating again the sensitivity of EDMs to
CP-violating parameters.

As mentioned above, also the constraint from $\tilde C_S$ is relevant: while in this case the constraint is numerically weaker, 
it is again at least a factor 100 stronger than an analogous constraint
from (semi-)leptonic processes in the A2HDM~\cite{Jung:2010ik,Celis:2012dk}.

For the neutron, the constraint induced in the charged-Higgs sector via the Weinberg operator is the dominant one. 
While this constraint does not imply sizable fine-tuning yet, it already prohibits large CP-violating effects in other observables in
specific models. For instance, while the indirect constraint from the branching ratio in $b\to s\gamma$ in the A2HDM still allows for a
sizable CP asymmetry for this process, a NP contribution of $|A_{\rm CP}(b\to s\gamma)|\lesssim1\%$ follows from the EDM bound and the
discussion in Refs.~\cite{Jung:2010ab,Jung:2012vu}.

These examples show the potential of EDMs, but also their complementarity to other searches, since only the imaginary parts of
parameter combinations are constrained. However, for the combinations EDMs are sensitive to, they are often the strongest constraints
available.

\section{cLFV}

Since lepton-flavour violation has been observed in neutrino oscillations, it should be present for charged leptons as well, even
within the SM. While the corresponding predictions are complicated by the fact that the neutrino sector is not fully specified, minimal
extensions yield typically tiny predictions for cLFV, $\sim \Delta m_\nu^2/M_W^2\sim 10^{-25}$ in the amplitude, due to the
GIM mechanism~\cite{Glashow:1970gm}, way below anything we can hope to detect in the foreseeable future.\footnote{Note that the fact
that with only one neutrino $\mu\to e\gamma$ would occur at a rate inconstistent with other weak transitions has been observed much
earlier~\cite{Feinberg:1958zzb}, although without proposing the GIM cancellation as the solution.}

Generic NP contributions can be many orders of magnitude larger, rendering cLFV observables excellent observatories for NP searches.
With a generic NP contribution suppressed by $1/M^2$, where $M$ is a mass scale characterizing the new contributions, the rate for cLFV
processes is simply suppressed by the square of this factor, $1/M^4$, implying that a given limit has to be improved by four orders of
magnitude in order to gain one order on the mass scale of NP. This is in contrast to the case of EDMs which constitute an
interference effect and therefore suffer only once from the strong suppression. 

As for EDMs, the low-energy description can be almost model-independently performed in the context of EFTs. However, similarly to EDMs
the analysis is complicated by the fact that the mass hierarchy is not necessarily the dominating one, given the involvement of various
small quantities. We concentrate here on the case of lepton-number conservation, \emph{i.e.} $\sum \Delta L_i=0$.
The leading $|\Delta L_i|=1$ operators are then radiative operators (mediating \emph{e.g.} $\mu\to e\gamma$), purely leptonic operators
(mediating \emph{e.g.} $\mu\to e\bar e e$), and semi-leptonic operators (mediating \emph{e.g.} $\mu\to e\bar q q$). At the electroweak
scale, there are furthermore operators containing Higgs- and heavy gauge-boson fields explicitly, discussed elsewhere at this
conference \cite{Crivellin:2016ekz,Brod}.
Experimentally, the three low-energy classes of operators have each their experimental equivalent in the sense that there are observables they contribute to on tree-level:
obviously $\ell\to \ell'\gamma$ and $\ell\to \ell'\bar{\ell'} \ell^{\prime\prime}$, and for $\ell\to \ell' \bar q q$  (semi-)leptonic
decays of mesons and $\mu\to e$ conversion in heavy atoms. On the loop-level, this simple correspondence ceases to exist.
Nevertheless, again similarly to EDMs, the hierarchy between these different classes and within different conversion processes can be
used to pin down the operators responsible once cLFV processes are observed, which in turn helps to identify the NP model responsible,
see \emph{e.g.} Refs.~\cite{Kitano:2002mt,Cirigliano:2009bz,Petrov:2013vka}. This observation implies that there is no single ``best'' measurement, but that as many
measurements as possible should be performed, for different transitions.

The existing experimental bounds are strongest for transitions between muons and electrons by far, at least numerically: most recently,
the MEG collaboration obtained the limit $BR(\mu^+\to e^+\gamma)\leq4.2\times 10^{-13}$ \cite{TheMEG:2016wtm}, while the limits on
$\mu\to e\bar e e$ and muon-electron conversion stem from the SINDRUM(II) collaborations. Significant improvements are expected from
ongoing and coming experiments, see the presentaions~\cite{Okada,Evans,Lancaster}. LFV processes involving the $\tau$ lepton are much
less constrained; while in some models large effects can be excluded from the bounds on $\mu\to e$ transitions, this clearly depends on the
specific flavour structure of the model. On the other hand, effects involving the $\tau$ could be enhanced, motivated by its
larger mass, its being part of the third generation, and also the presently observed $4\sigma$-hint of lepton-flavour non-universality
(LFNU) involving $b\to c\tau\nu$ transitions. In fact, LFNU generically also yields LFV, see \emph{e.g.}
Refs.~\cite{Glashow:2014iga,Bhattacharya:2014wla}, due to the rotation to the mass basis. This yields additional motivation to search
for LFV decays of $B$ mesons. However, ``typically'' does not mean ``necessarily''; exceptions have for instance been discussed in
Refs.~\cite{Celis:2015ara,Alonso:2015sja}.

\section{\label{sec::Conclusions} Conclusions}
EDMs and cLFV processes provide unique constraints for the CP- and lepton-flavour-violating sectors of NP models, respectively. A
potential discovery of any such process would be a major achievement, independent of its source.
The interpretation of bounds and potential measurements in terms of fundamental theory parameters requires the careful estimation of
theoretical uncertainties and is complicated by potential cancellations on various levels. While this problem can be addressed for the
EDMs of heavy paramagnetic systems to extract the electron EDM and scalar electron-nucleon coupling model-independently, a similar
approach including all relevant systems should be aimed at, but requires several additional measurements for different systems.

For the occuring combinations of parameters, EDMs typically provide the most stringent constraints. We demonstrated this explicitly for
general 2HDMs, and more specifically for the A2HDM, where large CP-violating effects in other observables are strongly bounded by the
existing EDM limits.

cLFV processes have a long history of excluding NP models, imposing very stringent bounds on any model that exhibits LFV. Again it is
crucial to search in as many channels as possible, as only the combination of many observables can yield information on the underlying
model. The recent hints at LFNU motivate additionally the search for LFV $B$ meson decays, however, there is no guarantee, since
models can exhibit LFNU without inducing LFV.

Given the present strength of these constraints, forthcoming experiments will test a crucial part of the parameter space and might turn
existing bounds into observations.

\bibliography{EDM_proceedings}
\end{document}